\documentclass[longbibliography,floatfix,reprint,prx,aps,twocolumn,superscriptaddress]{revtex4-2}
\usepackage[T1]{fontenc}
\usepackage{graphicx}
\usepackage{enumitem}
\usepackage{amsmath}

\usepackage{amsfonts}

\usepackage{amsmath}
\usepackage{amsfonts}
\usepackage{amssymb}
\usepackage{mwe}
\usepackage{graphicx}
\usepackage{color}
\usepackage{bm}
\usepackage{soul}
\usepackage[normalem]{ulem}

\usepackage[colorlinks,bookmarks=false,urlcolor=blue, citecolor=blue, linkcolor = blue]{hyperref}

\begin{document}
\title{High-temperature kinetic magnetism in triangular lattices}
\author{Ivan Morera}
\affiliation{Departament de F{\'i}sica Qu{\`a}ntica i Astrof{\'i}sica, Facultat de F{\'i}sica, Universitat de Barcelona, E-08028 Barcelona, Spain}
\affiliation{Institut de Ci{\`e}ncies del Cosmos, Universitat de Barcelona, ICCUB, Mart{\'i} i Franqu{\`e}s 1, E-08028 Barcelona, Spain}
\author{M\'arton Kan\'asz-Nagy}
\affiliation{Max-Planck-Institut f\"ur Quantenoptik, Hans-Kopfermann-Str. 1, 85748 Garching, Germany}
\affiliation{Munich Center for Quantum Science and Technology (MCQST), Schellingstr. 4, D-80799 M\"unchen, Germany}
\author{Tomasz Smolenski}
\affiliation{Institute for Quantum Electronics, ETH Z\"urich, CH-8093 Z\"urich, Switzerland}
\author{Livio Ciorciaro}
\affiliation{Institute for Quantum Electronics, ETH Z\"urich, CH-8093 Z\"urich, Switzerland}
\author{Ataç Imamoğlu}
\affiliation{Institute for Quantum Electronics, ETH Z\"urich, CH-8093 Z\"urich, Switzerland}
\author{Eugene Demler}
\affiliation{Institute for Theoretical Physics, ETH Zurich,
Wolfgang-Pauli-Str. 27, 8093 Zurich, Switzerland}

\begin{abstract}
We study kinetic magnetism for the Fermi-Hubbard models in triangular type lattices,
including a zigzag ladder, four- and six-legged triangular cylinders, and a full two-
dimensional triangular lattice. We focus on the regime of strong interactions, $U\gg t$
and filling factors around one electron per site. For temperatures well above the hopping
strength, the Curie-Weiss form of the magnetic susceptibility suggests effective
antiferromagnetic correlations for systems that are hole-doped with respect to $\nu=1$,
and ferromagnetic correlations for systems with electron dopings. We show that these
correlations arise from magnetic polaron dressing of charge carriers propagating in a
spin incoherent Mott insulator. Effective interactions corresponding to these correlations
can strongly exceed the magnetic super-exchange energy. In the case of hole doping,
antiferromagnetic polarons originate from kinetic frustration of individual holes in a
triangular lattice, whereas for electron doping, Nagaoka type ferromagnetic correlations
are induced by propagating doublons. These results provide a theoretical explanation of
recent experimental results in moire transition metaldichalcogenide materials. To understand many-body states
arising from antiferromagentic polarons at low temperatures, we
study hole-doped systems in finite magnetic fields. At low dopings and intermediate
magnetic fields, we find a magnetic polaron phase, separated from the fully polarized
state by a metamagnetic transition. With decreasing magnetic field, the system shows a
tendency to phase separate, with hole rich regions forming antiferromagnetic spinbags.
We demonstrate that direct observations of magnetic polarons in triangular lattices can
be achieved in experiments with ultracold atoms, which allow measurements of three
point hole-spin-spin correlations.
\end{abstract}
\maketitle

\textbf{Introduction.} 
Achieving electric control of magnetism is a long standing goal in condensed matter physics. Such systems will not only provide new insights into magnetism of itinerant electron systems but also
hold the promise of realizing new types of devices that combine long term robustness of magnetic memory with the
fast electric control~\cite{Awschalom2007,Wolf2001}. One of the most studied examples of electrically controlled magnetism comes from magnetic
semiconductors~\cite{Ohno2000,Chattopadhyay2001,Kaminski2002,Boukari2002,Jungwirth2006,Lee2009,Nishitani2010,Sawicki2010,Li2012,Wen2013,Matsukura2015}. In this class of materials conduction band electrons control interaction between the localized spins and makes it possible to modify the ferromagnetic Curie transition temperature by applying a gate voltage~\cite{Ahn2006,Matsukura2015}. Interplay of charge carrier dynamics and magnetism is also crucial for understanding rich phase diagram of Colossal Magneto Resistance manganite materials~\cite{Millis1996,Tokura2000,Salamon2001,Tokura2006}. In this paper we discuss a different mechanism for electrical control of magnetism, in which by changing charge carrier concentration one can tune between the antiferromagnetic and ferromagnetic interactions. Our results provide theoretical explanation of the recently observed magnetism in transition metaldichalcogenide (TMDc) moire materials, in which transition between ferro and antiferromagnetic interactions has been observed in triangular superlattices close to unity filling factor~\cite{Tang2020,exp}. One of the key results of our study is that one can observe appreciable magnetic interactions at temperatures
far exceeding the superexchange interaction energy $J$. We find that effective magnetic interaction arises from the
kinetic energy of charge carriers and the relevant energy scale is set by their inter-site tunneling/hopping. Our work is also related to the recent observation of ferromagnetism in the Wigner crystal states of electrons at temperatures exceeding the expected superexchange interactions~\cite{Hossain2020,Kim2022}.

Our work builds upon earlier studies by Nagaoka~\cite{Nagaoka1966} and Haerter and Shastry~\cite{HS2005} of magnetism in Fermi-Hubbard type
models in the limit of large interaction strength $U$. Nagaoka’s theorem states that on a square lattice the ground state of a single hole-doped Fermi-Hubbard model is a ferromagnetic state. Haerter and Shastry extended these arguments to triangular lattices and demonstrated that for the same conditions the ground state exhibits anti-ferromagnetic correlations. 
The key argument of our paper is based on considering Fermi-Hubbard model on triangular-type lattices (see Eq.~\eqref{Eq:Hubb} for the precise form of the Hamiltonian) in the regime when single particle tunneling $t$ is much smaller than the local Coulomb interaction $U$ and the filling factor is close to one. We focus on the regime of temperatures that is much higher than the superexchange energy $J = 4t^2
/U$, but comparable to the single particle tunneling $t$. This regime is relevant for the current experiments with TMDCs heterostructures. While naively one expects to find no magnetic interactions in this regime, we demonstrate that doping the system away from $\nu = 1$ introduces antiferro-and ferromagnetic interactions
for hole and electron dopings respectively. In the context of cold atoms in optical lattices, magnetic correlations induced by propagation of charge carriers at high temperature has been previously suggested for the square lattice~\cite{Marton2017}.

Our work highlights the importance of magnetic polarons in correlated Mott insulators~\cite{Bulaevskii1968,Trugman1988,Schmitt1988,Shraiman1988,Sachdev1989,Kane1989,Dagotto1989,Auerbach1998,Nagaev2001,Grusdt2018,Koepsell2019,Soriano2020}. Most of the previous studies of
magnetic polarons in the Fermi-Hubbard model focused on the square lattice. They are believed to play a crucial role in unusual properties of high-$T_c$ cuprates, including both the pseudogap regime and d-wave superconductivity. Magnetic polarons on triangular lattices have also been a subject of theoretical studies~\cite{Smirnov1980,Batista2017,Morera2021}. It is expected that they also play an important role in defining properties of moire materials. Their relevance for moire materials has been suggested previously in~\cite{Morera2021,Liang2022,Lee2022}.

\textbf{Emergence of magnetic interactions from kinetic frustration in triangular lattices.} 
To show the appearance of magnetic interactions from kinetic frustration at finite temperature we will consider spin-$\frac{1}{2}$ fermions described by a single-band Fermi-Hubbard model,
\begin{align}
    H &= -t \sum_{\langle i,j\rangle,\sigma } \left(c_{i,\sigma}^{\dagger}c_{j,\sigma} + \textrm{h.c.}  \right) + U \sum_{i}n_{i,\uparrow}n_{i,\downarrow} \label{Eq:Hubb}\\
    &-\frac{h}{2}\sum_{i}\left(n_{i,\uparrow} - n_{i,\downarrow}\right), \nonumber
\end{align}
where $U$ is the on-site repulsion, $t$ is the hopping strength and $h$ is an external magnetic field. We define the filling of the system $\nu=1+\epsilon$ and we study the magnetic phases at finite temperature while changing the doping of the system from a hole-doped regime ($\epsilon<0$) to a doublon doped one ($\epsilon>0$).
\begin{figure}[t!]
        \centering
        \includegraphics[width=1\columnwidth]{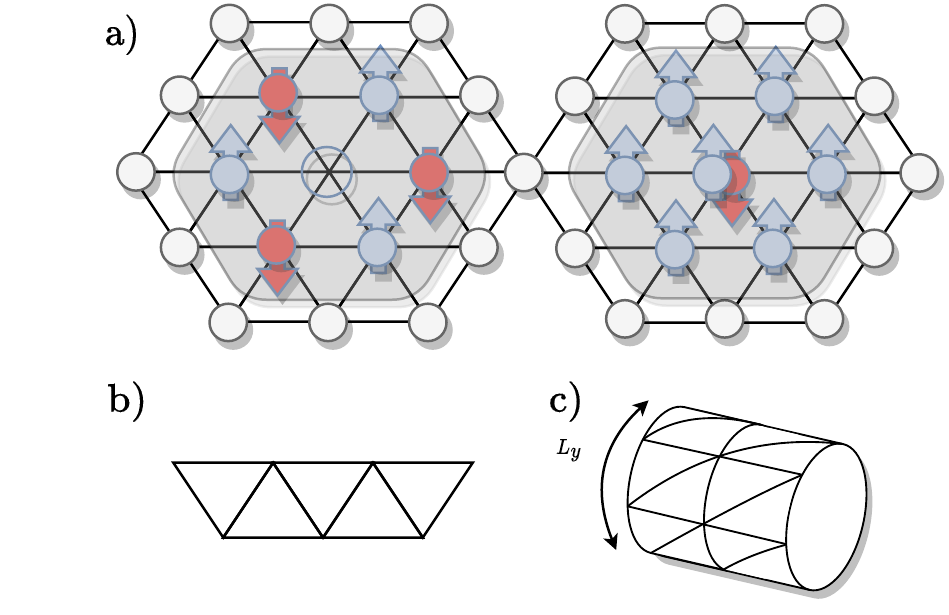}
        \caption{Triangular lattice geometries analyzed in this paper: 2d lattice (Monte Carlo), zigzag ladder and four- and six-legged triangular  cylinders (Tensor Network calculations), see panels b) and c). For the filling factor $\nu=1$  and at temperatures much larger than the superexchange interaction $k_BT>J$ we find a system of essentially decoupled spins. However when a single hole (doublon) is introduced on top of the spin incoherent Mott insulator antiferromagnetic (ferromagnetic) correlations appear surrounding it, see panel a). }
        \label{Fig:Sch}
\end{figure}

In the strongly interacting regime $U/t\rightarrow \infty$ for $\nu=1$ magnetic properties of the system are accurately described by a model of decoupled spins, that can be polarized by  a magnetic field of the order of temperature. However, if the system is doped away from unit filling the movement of charge carriers can induce different forms of magnetism. In a non-bipartite geometry the propagation of a single fermionic hole is frustrated in a polarized background~\cite{HS2005,Sposeti2014,Batista2017,Morera2021}. This can be seen from a high-$T$ expansion in the strongly interacting regime. Let us consider a spin polarized background with a single hole propagating through it. 
By fixing the initial and final points of the propagation we have two sets of paths characterized by the order $n$ of the high-$T$ expansion in $\beta t$. 
Since the fermionic hole has an effective negative hopping~\cite{HS2005,Sposeti2014,Batista2017,Morera2021} we see that odd paths contribute with an opposite sign than even paths. This gives a destructive interference pattern in the hole propagation. This concept is known as kinetic frustration since the hole cannot get the full kinetic energy associated with the geometry~\cite{HS2005,Sposeti2014}. However, if the background is not fully polarized then the destructive interference is suppressed when different hole trajectories result in distinguishable spin configurations. Thus kinetic energy of a hole can be lowered by inducing antiferromagnetic spin correlations around it. This effect is commonly referred to as antiferromagnetic correlations releasing kinetic frustration for a single hole. It underlies effective antiferromagnetic interactions in hole-doped Mott insulators in triangular type lattices. This effect is closely related to formation of a hole-magnon bound state in the case of spin polarized Mott insulators in triangular type lattices discussed previously in Refs.~\cite{Batista2017,Morera2021}. When we consider the electron doped regime ($\nu>1$) the propagation of doublons in the case of spin polarized background, is not frustrated since they effectively have a positive hopping. Furthermore, different doublon trajectories interfere constructively, in the case of ferromagnetic background, but not in the antiferromagnetic environment. In the latter case, different trajectories lead to distinguishable spin configurations. This induces an effective ferromagnetic interaction at ($\nu>1$)~\cite{Marton2017}. Notice that this transition between antiferromagnetic and ferromagnetic interactions by changing the doping at $\nu=1$ can only occur in a non-bipartite geometry. In a bipartite one there cannot be a destructive interference pattern since all paths contribute with the same sign in the high-$T$ expansion.

\begin{figure}[t!]
        \centering
        \includegraphics[width=1\columnwidth]{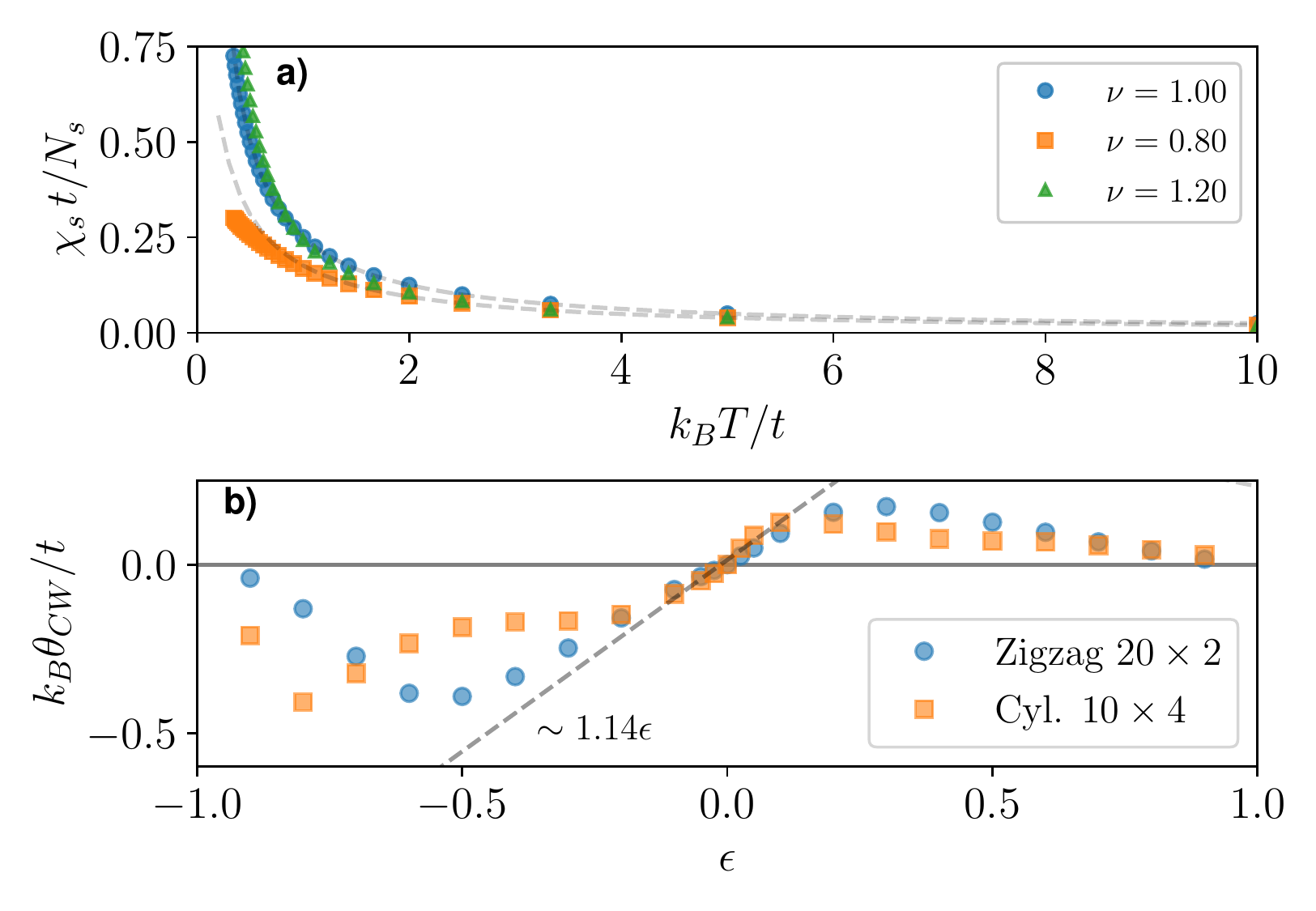}
        \caption{Panel a): Magnetic susceptibility per site as a function of temperature for three different dopings $\nu=1+\epsilon$ corresponding to a system at unit filling ($\epsilon=0$), a hole-doped system ($\epsilon<0$) and a doublon doped one ($\epsilon>0$) in a zigzag ladder $20\times 2$. Dashed lines correspond to fits using the Curie-Weiss law. Panel b): Critical temperature $\theta_{\textrm{CW}}$ as a function of doping $\nu=1+\epsilon$ for $U/t\rightarrow \infty$ and $h=0$. Circles correspond to a $20\times 2$ zigzag ladder and squares to a $10\times 4$ triangular cylinder. Dashed line indicates the linear dependence of the critical temperature with doping close to unit filling.}
        \label{Fig:CurWei}
\end{figure}
The appearance of kinetic magnetism has strong consequences in the magnetic properties of the system at finite temperature. When the typical temperatures are much higher than the superexchange interaction $J=4t^2/U$ we expect a paramagnetic phase at $\nu=1$. However, if the temperature is of the order of the hopping strength $k_BT \sim t$ we still expect to see the tendency to Haerter-Shastry antiferromagnetism in the hole-doped regime ($\nu<1$) and Nagaoka ferromagnetism in the electron doped one ($\nu>1$). In order to unravel the effects of kinetic magnetism at zero and finite temperature we will employ unbiased Tensor Network and high-temperature expansion quantum Monte Carlo simulations, see Fig.~\ref{Fig:Sch}. 
In a path integral picture, our Monte Carlo method samples the imaginary time paths of a hole or a doublon in a unit filling system with spin imbalance. The spin degrees of freedom are integrated out analytically, and only the paths are sampled. This leads to a low sampling noise for the spin correlations~\cite{Marton2017}, which allows us to achieve temperatures down to $k_B \, T\sim 0.2 \, t$ and system sizes up to several thousand sites. For details, see the Supplementary Material. 

\textbf{Magnetic susceptibility at finite temperature.}
The tendency towards magnetic order can be inferred from the magnetic susceptibility $\chi_s$ of the system. Upon doping we see an enhancement (suppression) of magnetic susceptibility above (below) unit filling at low temperatures, see Fig.~\ref{Fig:CurWei} panel a). This indicates the appearance of antiferromagnetic (ferromagnetic) correlations in the hole (electron) doped regime as expected. 
On the other hand, at large temperatures the magnetic susceptibility is suppressed in both regimes. This can be understood as holes and doublons are spinless particles that effectively reduce the possible total spin of the system thus reducing the magnetic susceptibility. 

\begin{figure}[t!]
        \centering
        \includegraphics[width=1\columnwidth]{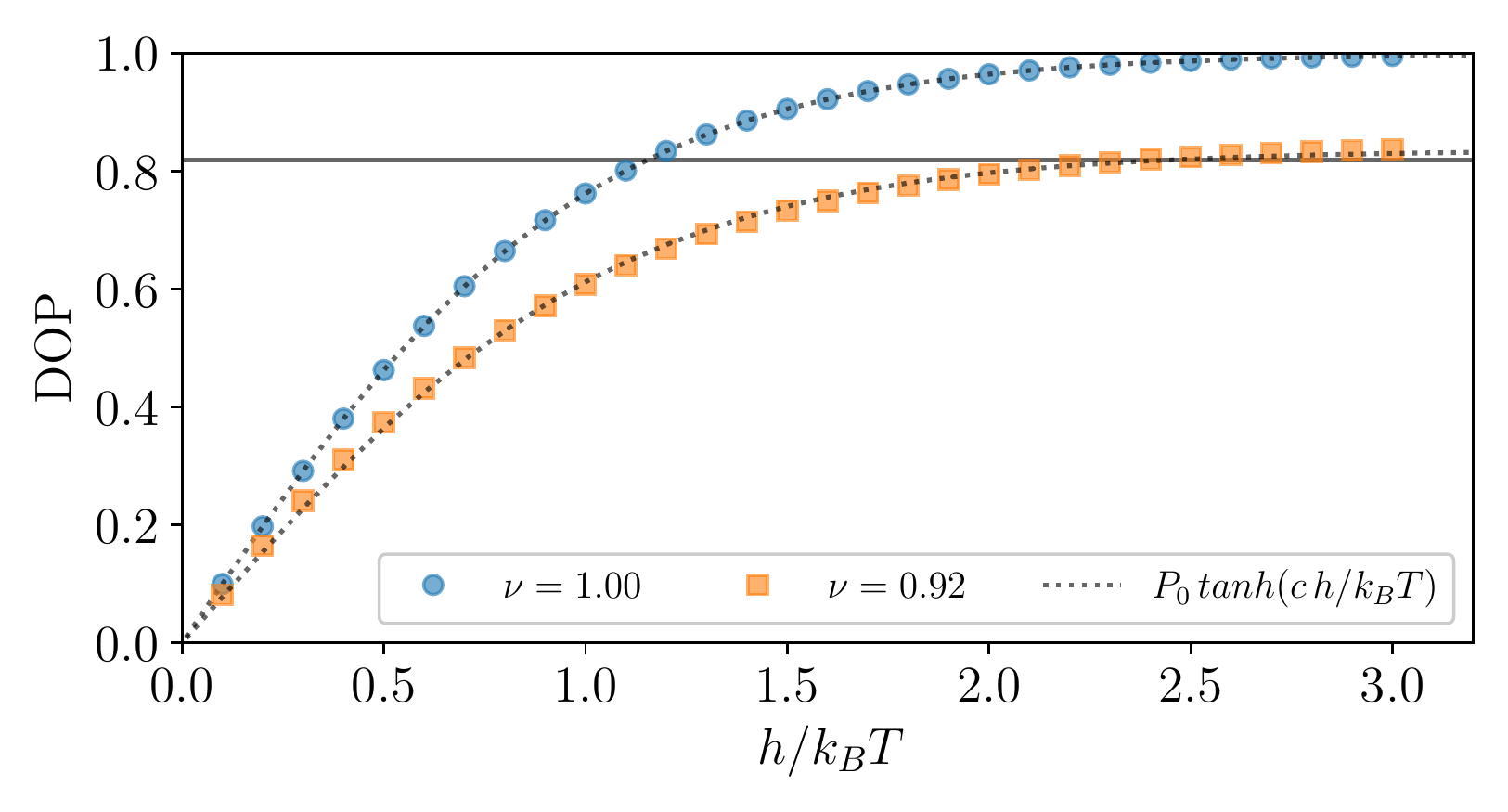}
        \caption{Degree of polarization (DOP) as a function of the magnetic field over temperature $h/k_BT$ at different dopings $\nu$ at a fixed ratio $h/t=1/4$ for a $6\times 6$ triangular cylinder. Dotted lines show the fit $P_0 \tanh(c h/k_B T)$. Continuous black line shows the polarization expected for a magnetic polaron gas, see main text.}
        \label{Fig:polT}
\end{figure}
In order to quantify the different contributions to the magnetic susceptibility we fit our data at intermediate temperatures ($10t>k_B T>t$) to a Curie-Weiss law, $\chi_s t /N_s = \frac{C}{T-\theta_{\textrm{CW}}}$,
where $C$ is the Curie constant and $\theta_{\textrm{CW}}$ is the critical temperature. The critical temperature denotes the tendency towards ferromagnetism ($\theta_{\textrm{CW}}>0$) or antiferromagnetism ($\theta_{\textrm{CW}}<0$). However at high temperatures, the constant $C$ dominates the dependence of the magnetic susceptibility with temperature. The Curie constant can be easily determined by taking into account the contribution of each spin to the spin susceptibility, $C=S(S+1)(1+|\epsilon|)/3$. Thus it decreases upon doping since the number of spins is reduced. On the other hand, the critical temperature changes sign at unit filling confirming the transition from an antiferromagnet to a ferromagnet, see Fig.~\ref{Fig:CurWei} panel b). We observe a linear tendency of the critical temperature with $\epsilon$ close to unit filling, $\theta_{\textrm{CW}}\sim 1.14 \epsilon t$, which confirms the appearance of kinetic magnetism. Experimentally, such behavior has been observed in moire materials in references~\cite{Tang2020,exp}.

\textbf{Nonlinear response to magnetic field.}
The linear magnetic response of the system at high temperature has allowed us to determine the tendency to antiferomagnetic (ferromagnetic) order in the hole (electron) doped regime. In the electron doped regime we expect a trivial ferromagnetic state at zero temperature. However the hole-doped regime presents a more exotic phase diagram as a function of doping. In order to elucidate the different many-body phases in this regime we study the nonlinear response to an external magnetic field. We define the degree of polarization (DOP) of the system as two times the magnetization per single fermion in the system $\textrm{DOP}=2m/(1-|\epsilon|)$.  
At unit filling $\nu=1$ the system follows a paramagnetic response as a function of the external magnetic field $\textrm{DOP}=\tanh(h/k_B T)$, see Fig~\ref{Fig:polT}. As the system is cooled down the magnetization increases until a critical temperature is reached at which the system fully polarizes. However upon hole doping the polarization is reduced. We observe that at fixed ratio $h/k_B T$ doping changes the polarization which points to the appearance of an effective magnetic interaction mediated by charge carriers. In order to quantify the effects of hole doping we fit our data to the generalized law $\textrm{DOP}=P_0\tanh(c h/k_B T)$, where we introduce the maximum polarization $P_0$ and the renormalization factor $c$. We observe that upon doping the system resists to polarize at weak fields $c<1$ which points to effective antiferromagnetic interactions. 
Moreover the system does not develop full polarization even at arbitrary small temperatures $P_0<1$. This indicates a phase transition to the formation of antiferromagnetic domains. In particular, for the chosen value of the magnetic field the system at small temperatures $k_B T < t$ is in a magnetic polaron phase as we will discuss below. In this phase each hole is attached with a spin flip (magnon) forming a hole-magnon bound state. Due to this effect the polarization in the magnetic polaron gas is linked to the hole density by, $P_0=(1-3|\epsilon|)/(1-|\epsilon|)$. 

As we have shown different many-body phases appear when doping the system at temperatures below the hopping strength $k_B T<t$. In the following we present the full response of the system to an external magnetic field for different hole dopings at zero temperature. This will allow us to show the different many-body phases that appear in the system due to kinetic frustration.
\begin{figure}[t!]
        \centering
        \includegraphics[width=1\columnwidth]{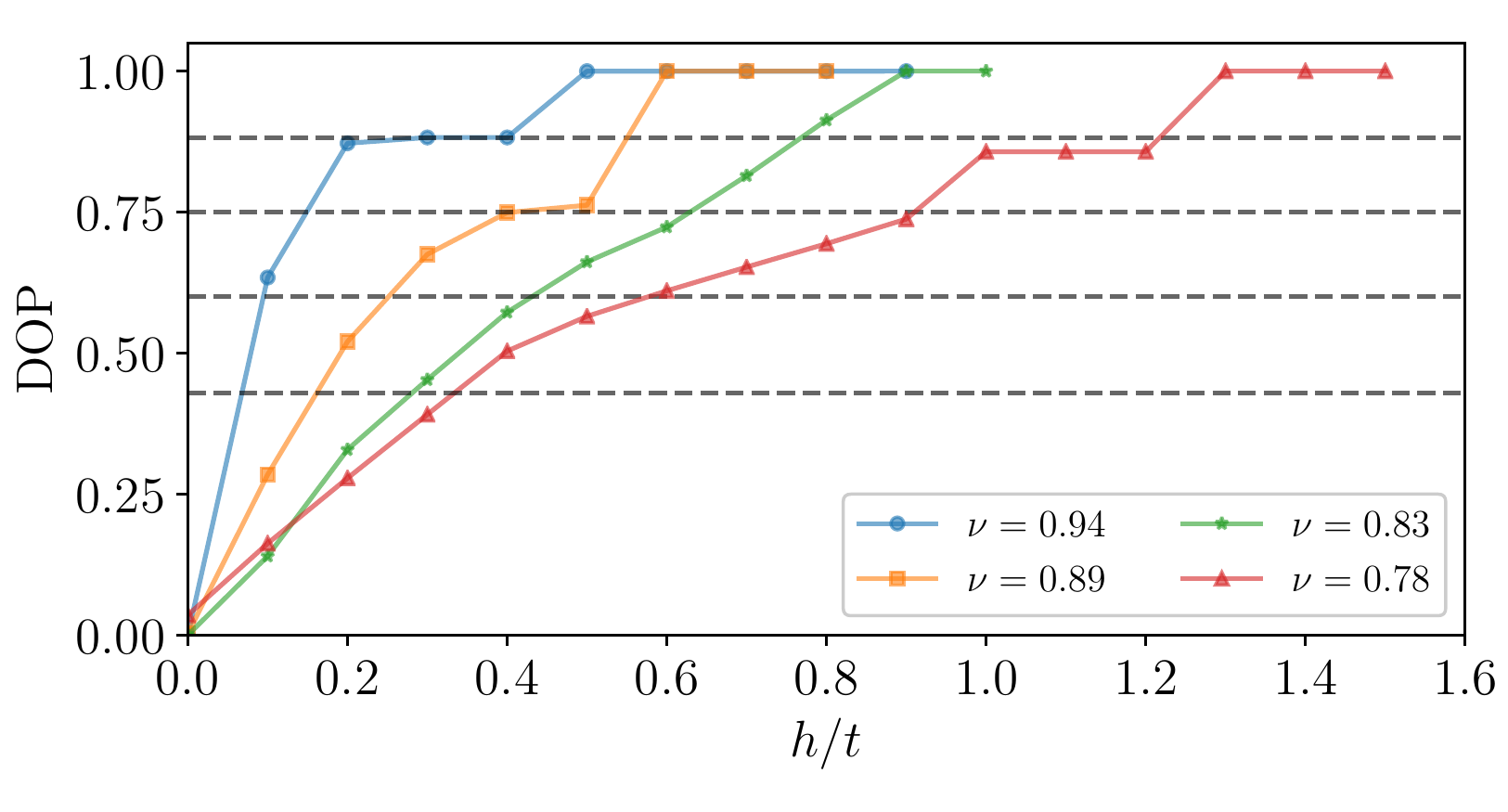}
        \caption{Degree of polarization (DOP) of an infinite triangular cylinder of $L_y=6$ as a function of the external magnetic field $h$ at different filling factors $\nu$ in the limit $U/t\rightarrow \infty$ and at zero temperature. Dashed lines indicate the polarization plateaus expected for the magnetic polaron gas at each filling factor, see main text.}
        \label{Fig:magfield}
\end{figure}

As shown in Fig.~\ref{Fig:magfield} the system shows a characteristic polarization versus magnetic field curve at zero temperature in the hole-doped regime. Upon hole doping the system resists to fully polarize even at zero temperature. We observe a large saturation magnetic field of the order of the hopping strength. At low dopings ($1>\nu\gtrsim 0.9$) the system has a sudden drop of polarization at a critical magnetic field. This effect resembles the one expected in metamagnetic materials. At this point the system has a first order phase transition from a fully polarized phase to a magnetic polaron gas. This phase is characterized by a fixed polarization and a gap to single particle excitations. Moreover the critical magnetic field at which the phase transition occurs is determined by the binding energy of the hole-magnon bound state $h_c/t\sim 0.5 t$. As the hole density increases the sharp signatures of the magnetic polaron gas start to blur. The polarization plateau is lost and the jump of polarization at the critical magnetic field disappears. This effect can be explained by taking the finite size of the magnetic polaron into account. At a critical hole density $\nu\sim 0.85$ magnetic polarons start to overlap due to their finite size. At this density we cannot rely on the simple picture of a weakly interacting gas of magnetic polarons. Instead of a sharp transition from a fully polarized state to a magnetic polaron gas we observe a smooth crossover in a wide range of magnetic field, see curves with $\nu<0.89$. In this region of magnetic field we expect a phase coexistence between magnetic polarons and bare holes. This phase coexistence is accompanied by a sudden change of slope in the degree of polarization as a function of magnetic field. The change of slope seems to appear when the system reaches the polarization of the magnetic polaron gas. 

At small magnetic fields the system linearly polarizes with the magnetic field, see Fig.~\ref{Fig:magfield}. In this regime the system phase separates into two domains with different hole densities. One domain has zero doping and is fully spin polarized, the other domain is "hole rich" and exhibits antiferromagnetic correlations. Appearance of phase separation indicates effective attractive interaction between magnetic polarons. While in the  "plain-vanilla" Hubbard model we find  global phase separation, we expect that introducing additional interactions, such as long range interactions, can result in other interesting phases, including the paired state and stripe order. The former may be relevant to superconductivity in moire systems. We point out the unusual feature of the effective interaction between magnetic polarons: it depends on the magnetic field. At small fields we find attractive interactions, resulting in phase separation. At larger fields, we find effective repulsion, leading to the magnetic polaron phase. To locate the spin bag phase in our system we determine the magnetic field at which multiple holes are confined inside a single bag. At this magnetic field we have a transition from a magnetic polaron gas to a spin bag phase at small dopings. As magnetic field is decreased, the size of the spin bag increases until it occupies the entire system at zero magnetic field. In this limit we recover Haerter-Shastry antiferromagnetism at low dopings.

\textbf{Implications for cold atoms experiments.}
\begin{figure}[t!]
        \centering
        \includegraphics[width=1\columnwidth]{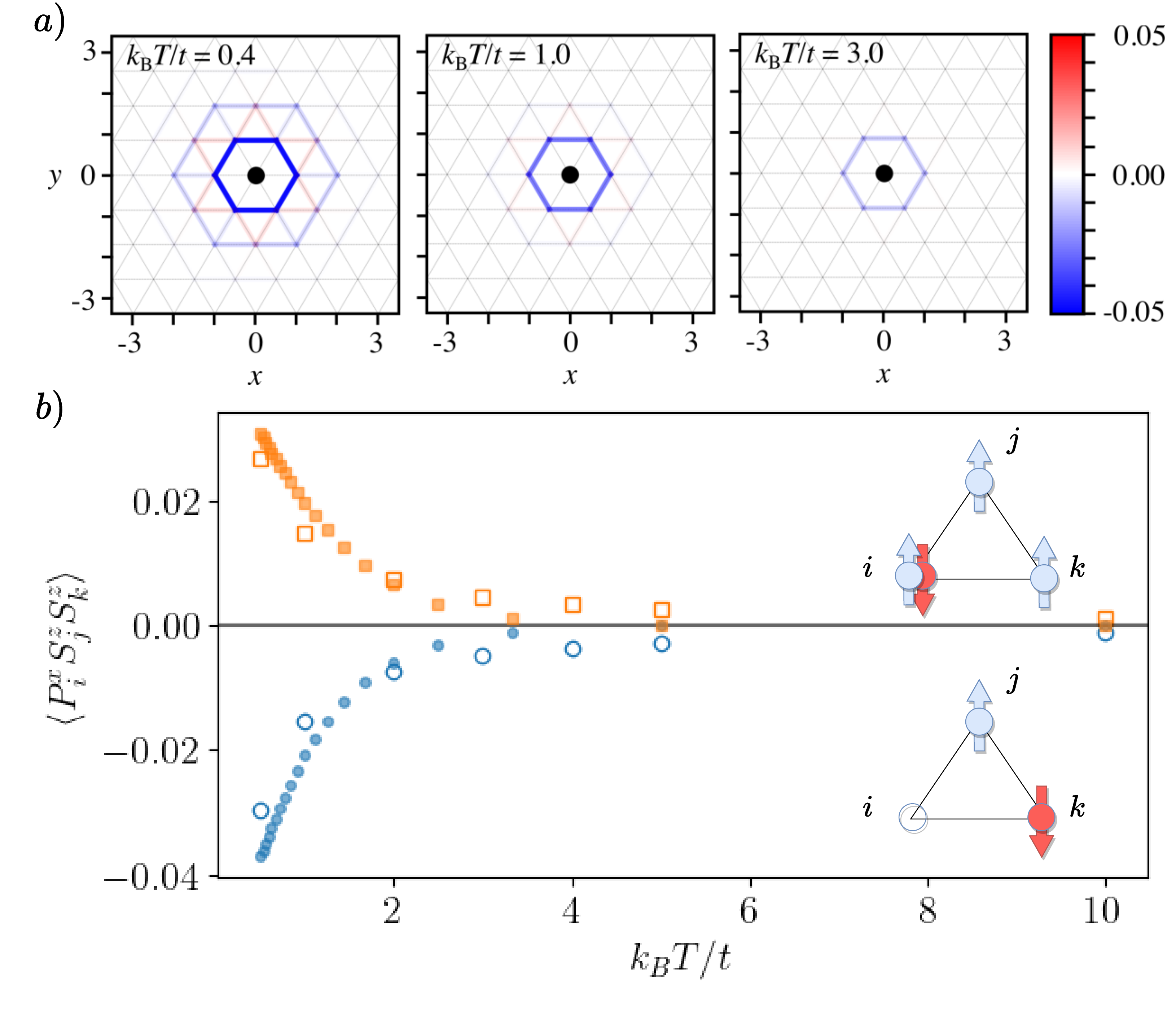}
        \caption{Panel a): Charge-spin-spin correlation function in a full two-dimensional triangular lattice $20\times 20$ obtained via Monte Carlo at three different temperatures. The black circle denotes the hole position and the links show the strength of the spin-spin correlations with colour. 
        Panel b): Charge-spin-spin correlation function as a function of temperature for a four-legged triangular cylinder with $L_x=8$ with a single hole (blue dots) and with a single doublon (orange squares). $P^x_i$ denotes a projector which fixes the charge position at site $i$. Open symbols correspond to the Monte Carlo results.
        We also show a schematic representation of the antiferro- and ferromagnetic correlations surrounding the hole and the doublon respectively.}
        \label{Fig:Correlations}
\end{figure}
Our results are not only relevant for TMDCs heterostructures but they also have strong implications for ultracold atomic systems in triangular-type optical lattices~\cite{Struck2011,Schauss}. Specifically, spin-charge correlations could be accessed in systems with a quantum gas microscope~\cite{Bakr2009,Sherson2010,PhysRevLett.114.213002,PhysRevLett.114.193001,PhysRevLett.115.263001,Haller2015,PhysRevA.92.063406,Greif953,Brown1385,Weitenberg2011,Schauss}. These experimental systems can be used to measure higher order correlation functions. In Fig~\ref{Fig:Correlations} we present the plaquette charge-spin-spin correlation function in the hole-doped regime ($\nu<1$) and the electron doped one ($\nu>1$) as a function of temperature. In the hole (electron) doped regime antiferromagnetic (ferromagnetic) correlations appear in the system as it is cooled down. Moreover the strength of the correlations is symmetric respect to unit filling. 
Thus we conclude that site resolved correlations could be used as a fingerprint of kinetic magnetism. Moreover charge-spin correlations will also denote the formation of magnetic polarons.

\textbf{Outlook.}
Our work shows how effective magnetic interactions arise due to the motion of charge carriers in a frustrated geometry at finite and zero temperatures. In the strongly interacting regime $U/t\rightarrow \infty$ the superexchange interaction is suppressed. However effective magnetic interactions due to motion of charge carriers are relevant when the system is doped away from unit filling. Due to kinetic frustration these effective magnetic interactions change sign close to unit filling. Thus a magnetic transition from an antiferromagnet to a ferromagnet is expected when moving from a hole-doped to an electron doped regime. We also point out the relationship of these effective magnetic interactions with the formation of magnetic polarons. We observe the formation of a magnetic polaron gas at temperatures below the hopping strength. 

Our results are relevant for recent experiments with TMDCs heterostructures which have observed a transition from antiferromagnetic to ferromagnetic interactions close to unit filling~\cite{Tang2020,exp}. Moreover the formation of magnetic polarons and the appearance of kinetic magnetism could be addressed in ultracold atoms laboratories through measurements of charge-spin correlations. Our work also highlights the occurrence of hole pairing through magnetic polaron binding. This could be relevant to explain the appearance of superconducting states near Mott insulators. In this paper we addressed the phase diagram in the regime of small hole doping. Additional phases can be expected at larger dopings. These will be addressed in the future publication. When this work was being finalized, we learned about the paper~\cite{Lee2022} that addressed related questions using a different theoretical technique.

\textbf{Acknowledgments.} 
The authors thank useful discussions with W. Bakr, A. Bohrdt, M. Greiner, F. Grusdt, E.-A. Kim, K. Sengstock, G. Refael, I. Esterlis.
I.M. thanks support from Grant No. PID2020-114626GBI00 from the MICIN/AEI/10.13039/501100011033 and Secretaria d'Universitats i Recerca del Departament d’Empresa
i Coneixement de la Generalitat de Catalunya, co-funded
by the European Union Regional Development Fund within
the ERDF Operational Program of Catalunya (project QuantumCat, Ref. 001-P-001644). I.M. acknowledges the Theoretical Physics
Institute at ETH for hospitality, where part of this work was completed. M.K.-N. acknowledges support by the EU Horizon 2020 program through the ERC Advanced Grant QUENOCOBA No. 742102 and from the DFG (German Research Foundation) under Germany’s Excellence Strategy – EXC-2111 – 390814868. 
E.D. acknowledges support from the ARO grant number W911NF-20-1-0163.
Tensor Network
computations have been performed 
using TeNPy~\cite{10.21468/SciPostPhysLectNotes.5}.

\bibliographystyle{apsrev4-2}
\bibliography{paperbib}

\pagebreak
\onecolumngrid
\vspace{\columnsep}
\newpage
\begin{center}
\textbf{\large Supplementary Material: High-temperature kinetic magnetism in triangular lattices}
\end{center}
\vspace{2cm}
\twocolumngrid

\setcounter{equation}{0}
\setcounter{figure}{0}
\setcounter{page}{1}
\makeatletter
\renewcommand{\theequation}{S\arabic{equation}}
\renewcommand{\bibnumfmt}[1]{[S#1]}
\addtolength{\textfloatsep}{5mm}

\section{Numerical details}
Due to the many competing orders in non-bipartite lattices we will employ unbiased Tensor Network simulations in two sets of frustrated geometries: a zigzag ladder and four- and six-legged triangular cylinders. We also employ Monte Carlo techniques to address the full two-dimensional limit.
We rely on unbiased numerical simulations due to the strong spin-charge coupling and the appearance of multi-body correlations which will not be captured by simple mean-field theories. Moreover our simulations allow to discern the different magnetic phases at finite and zero temperature. 

To study the finite temperature properties within Tensor Networks we construct a finite temperature ensamble using purification in finite systems in a canonical ensamble. We implement the $U(1)$ symmetry associated with charge conservation to fix the number of particles. However we do not fix the net magnetization. Starting from an infinite temperature state we progressively cool down the system by applying the Boltzmann factor $e^{-\beta H}$ to our MPS using the MPO $W_{\textrm{I}}$ technique. This one allows to take into account the long-range interactions induced in the quasi-1D systems~\cite{Zalatel2015}. The cooling process is performed using matrix product states of maximum bond dimension $\chi=512$. To measure the spin susceptibility we use fluctuation theorem $\chi_s k_B T N_s = \langle M^2\rangle - \langle M\rangle^2$, being $N_s$ the number of sites in the system. 
For studying the many-body phases at zero temperature we employ DMRG and iDMRG simulations. The DMRG simulations are obtained in systems with $x$ periodic boundary conditions to have faster convergence with system size. Our iDMRG simulations allow to access the properties of the system in the thermodynamic limit. To simulate the Fermi-Hubbard in the limit $U/t\rightarrow \infty$ we project out the Hilbert space: for $\nu<1$ ($\nu>1$) doublons (holes) are projected out. 

In order to tackle large two-dimensional systems, we also use a high-temperature expansion quantum Monte Carlo approach. We generalize the methods developed in Refs.~\cite{Carlstrom2016, Marton2017} for real-time dynamics of a single hole to the equilibrium case discussed in this work. This method allows one to consider arbitrarily large systems with an accuracy limited only by the number of Monte Carlo samples.  We consider a single hole or doublon in system of unit filling $\nu=1$ in the strongly interacting $U/t \to \infty$ limit. For simplicity, we only discuss the case of a hole. After a particle-hole transformation, the doublon case becomes identical to the hole case with the sign of the hopping flipped. Thus all formulas derived for the hole below apply to the doublons case with $t \to -t$.

At unit filling and strong interactions, the hole permutes the spins as it hops around the lattice. The Hamiltonian Eq.~\eqref{Eq:Hubb} simplifies to  $H_t = t\sum_{i^\prime i} h_{i^\prime}^\dagger \, h_i \, {\mathcal P}_{i^\prime i}$, where $h_i$ denotes the hole annihilation operator, $S_i^z = (n_{i\uparrow} - n_{i\downarrow})/2$ is the spin-$z$ operator at site $i$, and ${\mathcal P}_{i^\prime i}$ moves the spin from site $i^\prime$ to site $i$. 
In the absence of the hole, the correlation between any random spins would be $C_{bg} = \frac{1}{4}(p_\uparrow - p_\downarrow)^2$. These correlations are modified by the entanglement of the hole and the surrounding spins due to its imaginary time propagation, leading to the formation of a magnetic polaron.
The hole-spin-spin correlation function is given by
\begin{equation}
    C_{jl} = \frac{\sum_i {\rm Tr}_s\left( h_i S_{i+j}^z S_{i+l}^z \, e^{-\beta (H_t- h M)} \, h_i^\dagger\right) }{\sum_{i} {\rm Tr}_s\left(h_{i} \, e^{-\beta (H_t- h M)} \, h_{i}^\dagger\right)},
    \label{eq:correlation_function}
\end{equation} 
where ${\rm Tr}_s(\dots)$ denotes the trace over all possible spin states, and $M = \sum_{i^\prime}S_{i^\prime}^z$ is the total magnetization. Due to translational invariance, we can assume that the hole is always at the origin. Since the hopping $H_t$ conserves the total magnetization $M$, we can simply take the external magnetic field into account by assuming independent imbalanced spins at each site with probabilities of the up and down states given by $p_{\uparrow/\downarrow} = e^{\mp \beta h}/(e^{\beta h} + e^{-\beta h})$. 

We determine $C_{jl}$ by expanding the kinetic term $e^{-\beta H_t}$ in powers of $\beta = 1/(k_B T)$.
In the expansion, $\frac{(-\beta H_t)^n}{n!}$ corresponds to the contributions of all of the $z^n$ paths of length $n$, with $z=6$ denoting the coordination number of the lattice. Due to the sign of these terms, the hole picks up a phase factor $- {\rm sgn}(t)$ at each step, leading to both positive and negative contributions to the hole-spin-spin correlations in the frustrated triangular lattice, which results in an antiferromagnetic polaron cloud. In the doublon's case, however, the phase factor is always unity, hence the doublon's polaron cloud is ferromagnetic.

In the Monte Carlo procedure, we sample the paths randomly and sum their contributions both to the numerator and to the denominator of Eq.~\eqref{eq:correlation_function}. We draw the path length $n$ from a Poisson distribution $\mathbb{P}_n = \frac{(z \beta |t|)^n}{n!} \, e^{-z \beta |t|}$ to account for the absolute value $\frac{(z \beta |t|)^n}{n!}$ of the prefactors in the $n$-th order term $\frac{(-\beta H_t)^n}{n!}$~\cite{Carlstrom2016}. As compared to the high-temperature expansion approach which sums up all paths up to a given order $n_{max}$, the importance sampling of the Monte Carlo method allows one to reach much lower temperatures. Furthermore, it is numerically exact in the limit of infinitely many samples.

In order to determine the spin matrix elements of each path, we need to analyze its permutation effect on the spins.
Propagating over each path $(0, i_1, i_2, \dots, i_{n-1}, i_n)$ the hole permutes the initial state as $\mathcal{P}_{\rm path}  \, h^\dagger_{i_n} h_{i_0}= \mathcal{P}_{i_n i_{n-1}} \dots \mathcal{P}_{i_2 i_1}\mathcal{P}_{i_1 0} \, h^\dagger_{i_n} h_{0}$. 
Since only those states which are restored by this permutation lead to non-zero matrix elements, we only need to account for paths returning to the origin $i_n=0$, and for initial spin states which are invariant with respect to $\mathcal{P}_{\rm path}$. By decomposing $\mathcal{P}_{\rm path} = \prod_{k} \mathcal{C}_k$ as a product of irreducible permutation cycles $\mathcal{C}_k$, we can determine the fraction of invariant states among all random spin states. We can thus calculate the spin traces for each path exactly. In order to be unchanged by the permutation $\mathcal{C}_k$ all spins on sites in $\mathcal{C}_k$ have to be identical. The fraction of those spin states is
\begin{equation}
\alpha_{\rm path} 
= 
\frac{{\rm Tr}_s\left(h_{0} \, \mathcal{P}_{\rm path} \, e^{\beta h M} h_{0}^\dagger\right)}{\mathcal{Z}_s}
= 
\prod_k \left( p_\uparrow^{\#\mathcal{C}_k} + p_\downarrow^{\#\mathcal{C}_k} \right),
\nonumber
\end{equation}
where $\# \mathcal{C}_k$ denotes the number of spins in the cycle, and introduced the spin partition function $\mathcal{Z}_s = {\rm Tr}_s\left(h_{0} \, e^{\beta h M} h_{0}^\dagger\right)$ for normalization.
The contribution of each sampled path of length $n$ to the denominator in Eq.~\eqref{eq:correlation_function}, is thus $(-{\rm sgn}(t))^n \, \alpha_{\rm path}$. 

Similarly, we can determine the contribution of each path to the numerator $(-{\rm sgn}(t))^n \, \gamma_{\rm path}$, where
\begin{equation}
\gamma_{{\rm path}, j l} 
= 
\frac{{\rm Tr}_s\left(h_{0} \, S_{j}^z  S_{l}^z \, \mathcal{P}_{\rm path} \, e^{\beta h M} \, h_{0}^\dagger\right)}{\mathcal{Z}_s},
\nonumber
\end{equation}
 depends on the effect of $\mathcal{P}_{\rm path}$ on sites $j$ and $l$. The value of $\gamma_{\rm path}$ depends on whether the sites $j$ and $l$ are affected by the permutations $\mathcal{C}_k$. If $j$ and $l$ are not part of any of the permutation cycles, they are not permuted by the hole, thus
\begin{equation}
\gamma_{{\rm path}, j l} = \frac{\alpha_{\rm path} }{4} (p_\uparrow - p_\downarrow)^2= \alpha_{\rm path} C_{bg}.
\end{equation}
If only one of the spins $j$ is part of a cycle, $\mathcal{C}_j$, then it has be identical to other spins in $\mathcal{C}_j$. Since the probability of all spins being up/down in this cycle is $p_{\uparrow / \downarrow}^{\# C_j}$, 
\begin{equation}
\gamma_{{\rm path}, j l} = \frac{\alpha_{\rm path}}{4} \, (p_\uparrow - p_\downarrow) \, \frac{p_\uparrow^{\# \mathcal{C}_j} - p_\downarrow^{\# \mathcal{C}_j}}{p_\uparrow^{\# \mathcal{C}_j} + p_\downarrow^{\# \mathcal{C}_j}}.
\nonumber
\end{equation}
Similarly, if $j$ and $l$ are parts of different permutation cycles, $\mathcal{C}_j \neq \mathcal{C}_l$, then 
\begin{equation}
\gamma_{{\rm path}, j l} =\frac{\alpha_{{\rm path}}}{4} \; \frac{p_\uparrow^{\# \mathcal{C}_j} - p_\downarrow^{\# \mathcal{C}_j}}{p_\uparrow^{\# \mathcal{C}_j} + p_\downarrow^{\# \mathcal{C}_j}} \; 
\frac{p_\uparrow^{\# \mathcal{C}_l} - p_\downarrow^{\# \mathcal{C}_l}}{p_\uparrow^{\# \mathcal{C}_l} + p_\downarrow^{\# \mathcal{C}_l}}.
\nonumber
\end{equation} 
Finally, when $j$ and $l$ are part of the same permutation cycle, then their spins always need to be identical, hence the correlation $S_i^z S_j^z$ in these states is always $\frac{1}{4}$, and we get
\begin{equation}
\gamma_{jl, {\rm path}} = \frac{\alpha_{{\rm path}}}{4}.
\nonumber
\end{equation}

In the hole's case, the negative phase contributions from the paths with odd lengths lead to the Haerter-Shastry anitferromagnetic polaron, as a manifestation of the frustration in the lattice~\cite{HS2005}. This is in contrast to bipartite lattices, where all returning paths are of even length, hence the hole-spin-spin correlation function only has positive contributions, leading to Nagaoka ferromagnetism~\cite{Nagaoka1966}.

\begin{figure}[t!]
\centering
\includegraphics[width=1\columnwidth]{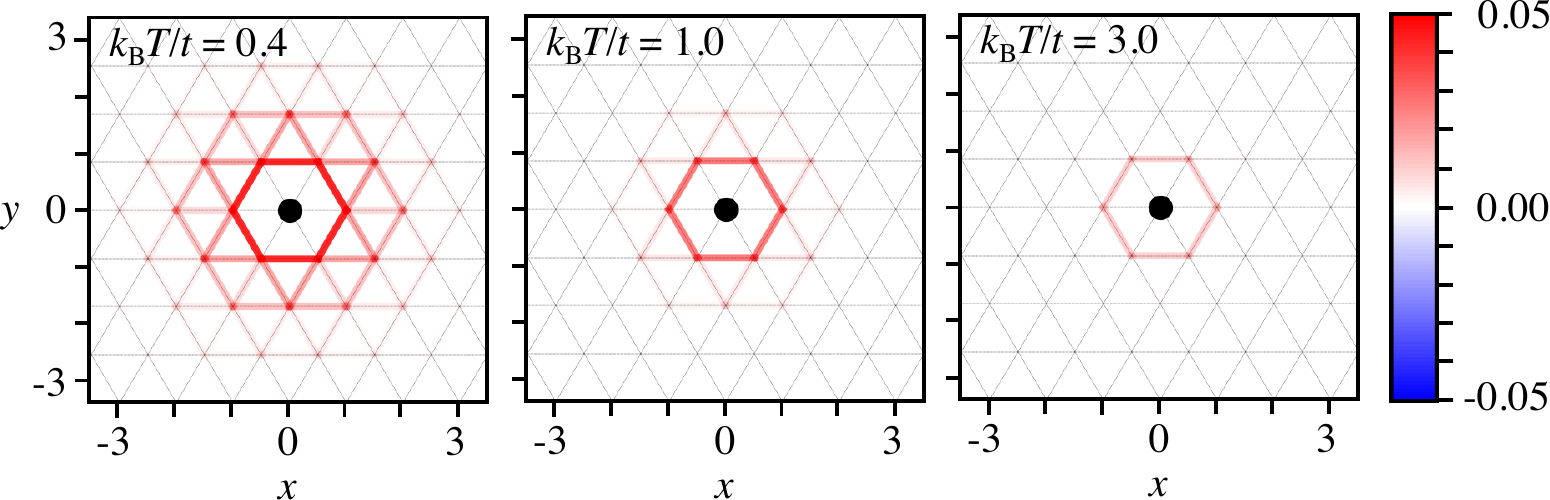}
\caption{Nearest-neighbor doublon-spin-spin correlation function in a two-dimensional $20 \times 20$ triangular lattice  at temperatures $k_B T/t= 0.4 $, $1$, and $3$. Similarly to the hole's case in Fig.~\ref{Fig:Correlations}~(a), the size of the magnetic bubble around the hole extends rapidly as the temperature is lowered.
The black circle denotes the hole position and the links show the strength of the correlations between the hole and the sites at the ends of the bond. Results were obtained using a Monte Carlo simulation.} 
\label{Fig:DoublonNNCorrelations}
\end{figure}

Doublon-spin-spin correlation functions can be determined using the same calculation but with a negative hopping. Due to this sign flip, all contributions to Eq.~\eqref{eq:correlation_function} will be positive. All doublon-spin-spin correlations will thus be ferromagnetic, as we demonstrate in Fig.~\ref{Fig:DoublonNNCorrelations}.

\section{Effective magnetic interactions}
As shown in the main text the critical temperature $\theta_{\textrm{CW}}$ depends linearly with the doping $\epsilon$ close to unit filling. This suggest the appearance of effective magnetic interactions due to charge motion $J_{\textrm{eff}}\sim -1.14 \epsilon t$. 
Moreover the same linear tendency is observed in the zigzag ladder and the triangular cylinders, see Fig. 1. in the main text. We observe deviations from the linear tendency at larger dopings in all geometries. The deviations seem to be enhanced in the triangular cylinders. In the electron doped regime the critical temperature saturates and starts to decrease approaching zero smoothly. In the hole-doped regime the critical temperature still decreases up to dopings $\epsilon\sim -0.8$ where it reaches a minimum and then increases up to zero.

In the strongly interacting regime $U\gg t$ where a $t-J$ model is a good description of the Fermi-Hubbard model Eq.~\eqref{Eq:Hubb} we expect a shift of the critical temperature $\theta_{\textrm{CW}}$ given by the antiferromagnetic superexchange interaction $J=4t^2/U$. Thus the effective magnetic interaction has two contributions $J_{\textrm{eff}}\sim J-1.14 \epsilon t$. In this situation the antiferromagnetic to ferromagnetic transition is shifted to the electron doped regime $\epsilon_c \sim 3.51 t/U$. Thus a finite density of doublons is needed to cancel out the intrinsic antiferromagnetic superexchange interaction. Thus we conclude that at high temperatures $k_B T>t$ the magnetic susceptibility can be expressed as,
\begin{equation}
\chi_s t/N_s \sim \frac{1}{3}S(S+1) \frac{1-|\epsilon|}{k_B T/t + 4t/U-1.14\epsilon}.
\end{equation}



\section{Magnetic polaron gas}
\begin{figure}[t!]
        \centering
        \includegraphics[width=1\columnwidth]{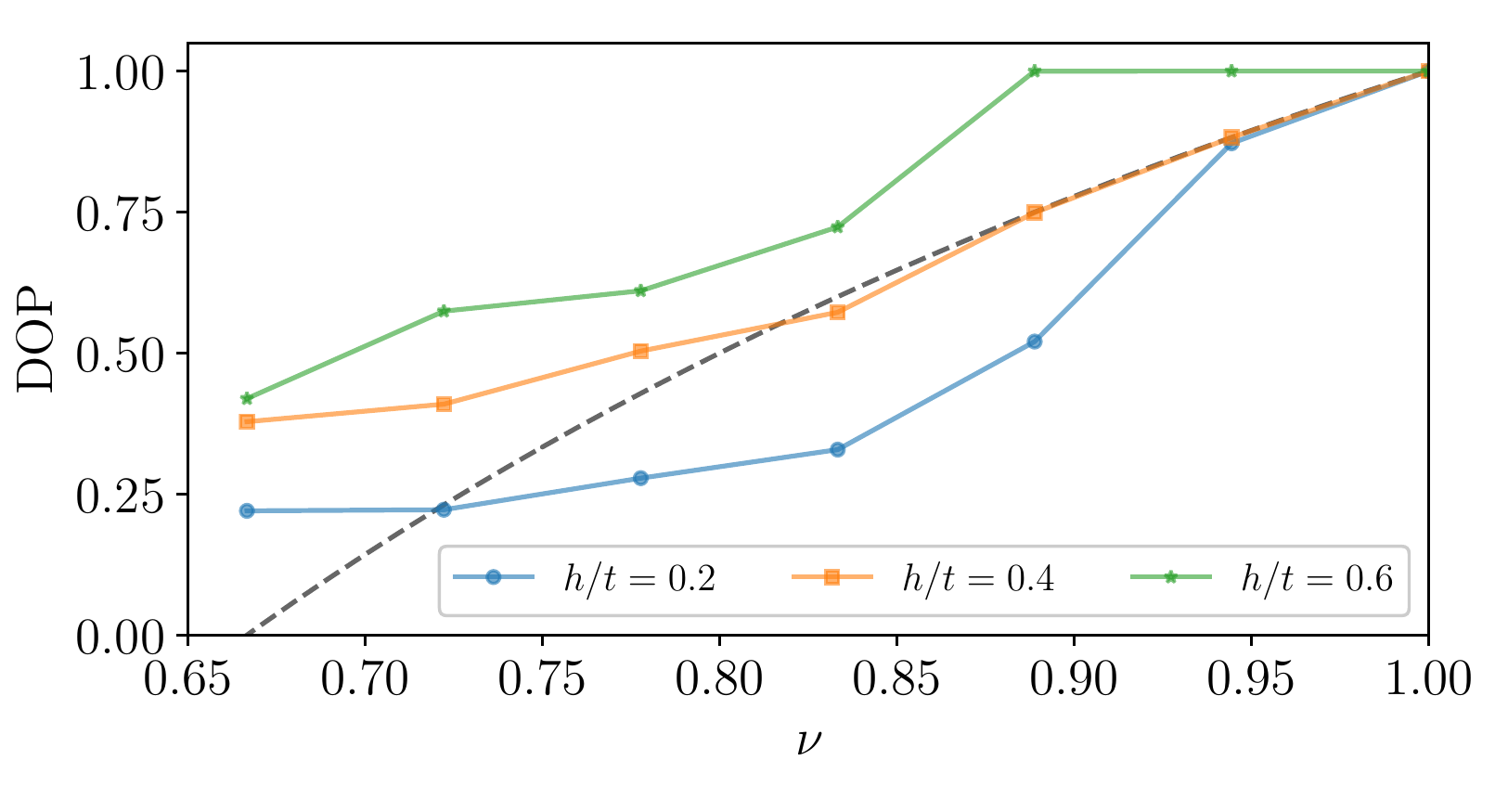}
        \caption{Degree of polarization (DOP) of an infinite triangular cylinder of $L_y=6$ as a function of the filling factor $\nu$ at different magnetic fields $h/t$ in the limit $U/t\rightarrow \infty$ and at zero temperature. Dashed line indicates the degree of polarization expected for the non-interacting magnetic polaron gas, $P_0=(1-3|\epsilon|)/(1-|\epsilon|)$.}
        \label{Fig:polarization_nu}
\end{figure}
The magnetic polaron gas is characterized by polarization plateaus as a function of magnetic field. This polarization is uniquely determined by the density of holes in the system $P_0=(1-3|\epsilon|)/(1-|\epsilon|)$.  
We present how the degree of polarization changes with hole doping at fixed magnetic field, see Fig.~\ref{Fig:polarization_nu}. We observe a general tendency of the system to reduce polarization as the density of holes increases. This feature agrees with the appearance of antiferromagnetic correlations proportional to the hole density. In the range of magnetic field $h/t\in [0.2,0.4]$ the system is in a magnetic polaron gas phase for small dopings $\nu>0.85$. For larger dopings in this range of magnetic field the polarization seems to saturate with doping. 
At large magnetic fields $h/t>0.4$ the system fully polarizes at small hole dopings and the polarization sudden decreases at a finite hole doping.

\section{Phase separation into spinbags}
\begin{figure}[t!]
        \centering
        \includegraphics[width=0.9\columnwidth]{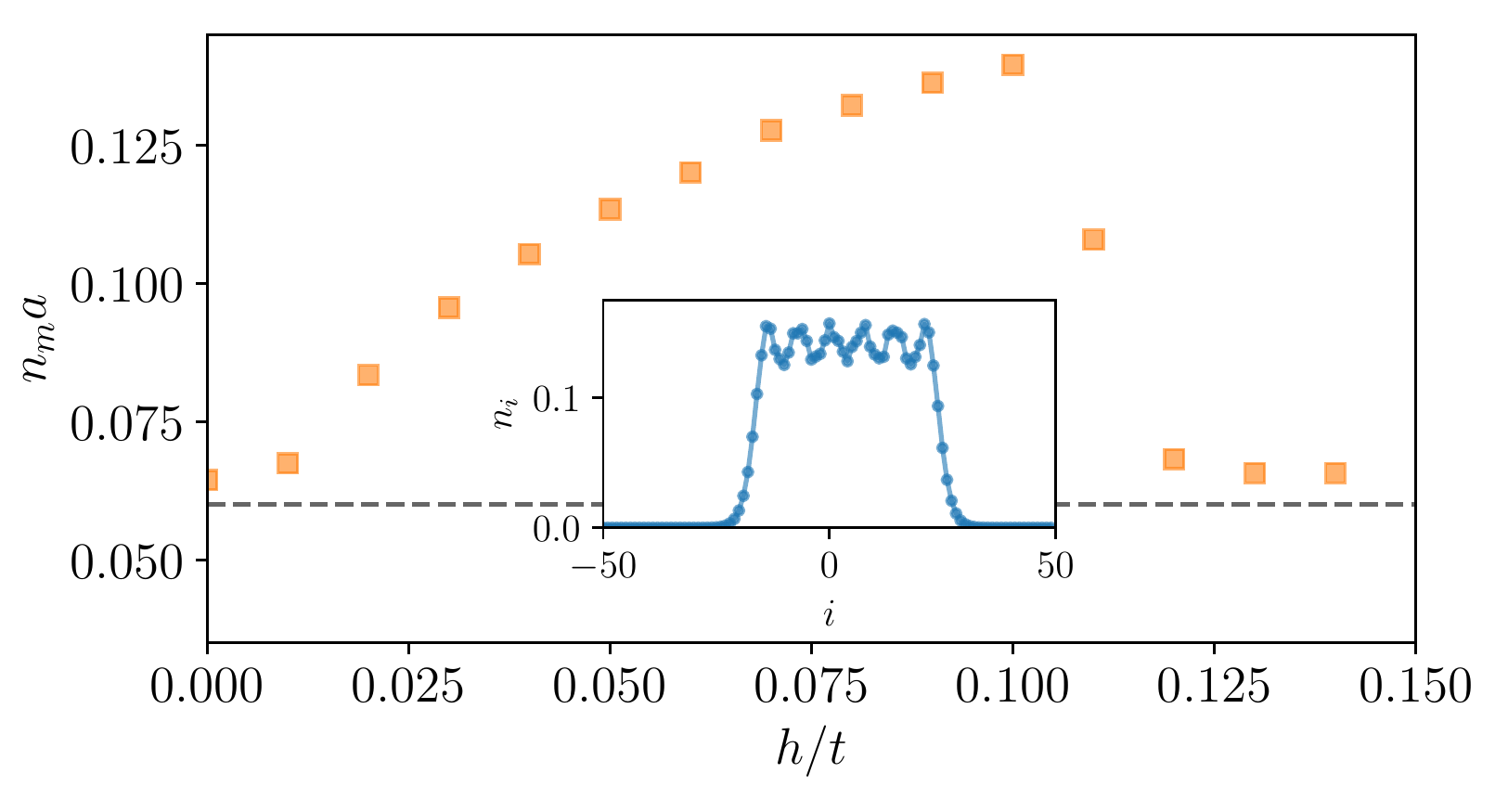}
        \caption{Main panel: Optimal hole density $n_ma$ of the antiferromagnetic spin bag as a function of the external magnetic field $h/t$ at $U/t \rightarrow \infty$ for a zigzag ladder with $50\times 2$ sites and $N_h=6$ holes. Dashed line indicates the mean density of the deconfined phase. Inset panel: Hole number as a function of the lattice index at $h/t=0.1$ for the same system as in the main panel.}
        \label{Fig:SpinBag}
\end{figure}

At small magnetic fields we expect the following structure: surrounding each hole an antiferromagnetic spin domain is formed and far away from it a ferromagnetic order is expected since the superexchange coupling is very small and the system can be easily polarized. This supports the idea of antiferromagnetic spin bags (ASBs). Holes are effectively confined inside these bags since they release their kinetic frustration by moving through an antiferromagnetic background, see inset panel of Fig.~\ref{Fig:SpinBag}. When the magnetic field is very small the size of the bags becomes very large and the entire system can form an antiferromagnetic order with a few number of holes. When multiple ASBs start to overlap then holes effectively deconfine since they can propagate through larger regions. As the magnetic field is increased the size of the bags is reduced and the magnetization increases. Thus a larger number of holes is needed in order to create an antiferromagnetic order through the entire system. 

The ASB has associated an optimum density of holes $n_m$ which can be related to their localization length, see Fig.~\ref{Fig:SpinBag}. For arbitrary small magnetic fields the bag is very large and this density drops to zero. For increasing magnetic fields the bag shrinks and the hole density increases. By further increasing the magnetic field the bag becomes so small that only a small number of holes can be accommodated within it. Thus the hole density starts to decrease again.  
At a critical magnetic field $h/t\sim 0.12$ the ASBs are so small that only a single hole can fit inside a single bag. At this point a transition to a magnetic polaron gas occurs at small hole dopings.

\end{document}